\documentstyle[12pt]{article}

\begin{document}

\begin{center}

{\large \bf Gravitational waves in an expanding Universe} \\

\vspace{2cm}

J\'ulio C\'esar Fabris\footnote{e-mail:fabris@cce.ufes.br}\\ 
\vspace{0.4cm}
{\it Departamento de F\'{\i}sica, Universidade Federal do
Esp\'{\i}rito Santo,} \\
\vspace{0.1cm}
{\it Vit\'oria CEP 29060-900-Esp\'{\i}rito Santo. Brazil.} \\
and\\
Sergio Vitorino de
Borba Gon\c{c}alves\footnote{e-mail:svbg@if.uff.br}\\
\vspace{0.4cm}
{\it Instituto de F\'{\i}sica, Universidade Federal
Fluminense}\\
\vspace{0.1cm}
{\it Niter\'oi CEP 24210-340-Rio de Janeiro. Brazil}
\vspace{1cm}

\end{center}
\centerline{\bf abstract}
We study the tensorial modes of the two-fluid model, where one of this
fluids has an equation of state $p = - \rho/3$ 
(variable cosmological constant, cosmic string fluid,
texture) or $p = - \rho$ (cosmological constant), while the other fluid is 
an ordinary matter
(radiation, stiff matter, incoherent matter). In the first case,
it is possible to have a closed Universe
whose dynamics can be that of an open Universe providing alternative
solutions for the age and horizon problems. This study of the
gravitational
waves is extended for all values of the effective curvature
$k_{eff}=k-\frac{8\pi
G}{3}\rho_{0s}$, that is, positive, negative or
zero, $k$ being the curvature of the spacelike section.
In the second case, we restrict ourselves to a flat spatial section.
The behaviour of gravitational waves have, in each case, very particular
features, that can be reflected in the anisotropy spectrum of
Cosmic Microwave Background Radiation. We make also some considerations
of these models as candidate to dark matter models.
\vspace{1.0cm}
\par
keywords: cosmology, large-scale structure of Universe.
\vspace{2cm}
\section{Introduction}
Some of the main problems today in cosmology are the determination of the
mass
parameter $\Omega$, the age of the Universe, and a consistent explanation
of
the thermal equilibrium in very early era
\cite{narlikar,ellis,peebles}. The mass parameter measures the ratio
between the total mass of
the Universe and the critical mass of a spatially flat Universe. The
inflationary model
predicts $\Omega = 1$, i.e., a spatially flat Universe. But
apparently the observational
status of the mass paremeter of the Universe leads to 
contradictory results: The luminous mass is about
$\Omega_L \sim 0.005$, while consistency with primordial nucleosynthesis
suggests a baryonic mass such that $\Omega_B \sim 0.02$; however,
the dynamic of motion of galaxies in a cluster, indicates a clustered mass
of other $\Omega_C = 0.3$ (the indices $L$, $B$ and $C$ standing for 
"luminous", "baryonic" and "clustered" repectivelly); on the other
hand, the position of the first doppler peak in the spectrum of anisotropy
of cosmic
microwave background radiation, which is generally fixed by the inverse of
the
square root of the total mass, indicates $\Omega_T \sim 1$, but the error
bar
is very large. Besides that, the
deviation of the Hubble law from the linearity is consistent with a model
with
cosmological constant of other $\Omega_\Lambda \sim 0.7$ and $\Omega_T =
1$, 
leading to an accelerating
Universe\cite{perlmutter,riess}. 
If this is the case, we could be living now in a phase dominated by the
cosmological constant.
\par
The age problem is linked
with a precise measurement of the Hubble parameter,
which is yet a point of controversy,
since some results point to an age for the Universe very near 
the measured age of globular cluster. The age of globular clusters is
estimated to be of the order of $t_c \sim 15 Gy$.
If the Universe is now in matter dominated phase,
so that $\Omega_M = 1$, the Einstein's equations imply that the
scale factor behaves as
$a \propto t^{2/3}$, and it results $H_0t_0 = \frac{2}{3}$,
where $H_0 = \frac{\dot a_0}{a_0}$ is the Hubble parameter today and
$t_0$ is the age of the Universe. Taking
$H_0 \sim 70 Km/s/Mpc$, we obtain $t_0 \sim 12 Gy$, that is, lesser
than the age of the globular clusters. However, there is indication
that better estimations of the distances can lead to smaller
age for the globular clusters and a bigger age for the Universe.
A more sophisticated method, taking into acount the velocity expansion,
and the possible existence of a cosmological constant gives
an estimation for the age of the Universe of order $t_0 =
15 Gy$, while a more precise evaluation of distances of astronomical
objects lead to an new estimation of the age of globular clusters
to be about $t_{GC} \sim 11.5\pm2Gy$. But to our knowledge, these
new estimations are not yet a consensus.
\par
Finally, the thermal equilibrium of the Universe
in its first moments is explained by inflating
a small causally connected region
to scales comparable with our observed Universe,
in the so-called inflationary period
\cite{guth}. Such mechanism seems necessary since
otherwise, it could be difficult to understand why we observe
the same temperature in regions that, at the time of emission of
the photons that we receive now, was not in contact.
The inflation can give an explanation for the thermal equilibrium,
but there is not
still at this moment an unique scenario.
\par
All these problems may be also treated by the
inclusion of non ordinary matter in the Universe.
In doing so, we modify the dynamics of the Universe,
consequently changing the estimation
for the clustered mass and age of the Universe.
In some specific cases, an
alternative
explanation for the isotropy of the Universe can be implemented.
Very employed in these last times are the so called cold dark matter
model (the effective pressure of the dark matter is zero) or
hot dark matter model (the pressure is that of a radiative fluid)
\cite{dark}.
Observations seems to favor of the cold dark matter model.
A more recent example is the so called "quintessence", a fluid component
that will be present in the Universe besides the ordinary fluid.
Its presence leads to an equation of state that varies from
a positive (or null value) to a negative one. One realization
of "quitenssence" is a scalar field $Q$ in a slowly decreasing
potential $V(Q)$. The quintessence has good consequences for the
age of the Universe and leads to a spectrum of perturbations
consistent with the observational data\cite{quinte1,quinte2}.
\par
In this article, we consider two main possibilities
of non ordinary fluid: a stringlike fluid
characterized by an equation of state $p = - \frac{\rho}{3}$,
and a cosmological constant. In both cases, we consider these "extra" 
matters coupled to ordinary one, that is, matter characterized by
an equation
of state of the type $p = \alpha\rho$, with $\alpha = 0, \frac{1}{3}$
and $1$.
The equation of state $p = - \frac{\rho}{3}$
characterizes the limiting region from which the strong energy condition
is
no longer satisfied
and where inflation takes place. The
energy density associated with this fluid decreases as
$\rho_s \propto a^{-2}$, where $a$ is the scale factor of
the Universe. Some kind of fundamental fields can be represented in
some sense by
such an equation of state:
variable cosmological constant\cite{ioav}; cosmic
string\cite{prl}; texture\cite{davis}. 
On the other hand, the reason to consider a model with
a cosmological constant term is evident from the considerations
made above.
\par
In both cases described above, there will be a period where
the "extra" fluid dominates
over the ordinary matter, so that the equation of state of
matter evolves progressively from a positive (null) value to
a negative one.
We can live now in a Universe already dominated
by this extra fluid.
This can lead
to an older Universe with respect to the Standard Cosmological Scenario,
avoiding the contradictions between the age of the Universe evaluated from
the Hubble law and the age of the globular clusters, for example.
\par
In the case of a stringlike fluid, there is another nice feature.
Such a fluid mimics a
curvature term in the Einstein's equation: the
topology of the space can be, for example, that of a closed
Universe, whereas the dynamics is that of an open Universe. This can solve
the horizon problem without inflation.
In reference \cite{prl} the confrontation of specific models with
observation leaded to some viable scenarios.
\par
Our main interest will be concentraded in the evolution of gravitational 
waves in a background Universe whose matter content is one of the
two fluid models described above.
We will
determine the solutions for an isotropic homogenous Universe. Then,
we will analyse the evolution of gravitational waves in such Universe.
One advantage to treat gravitational waves is that it is quite
sensible to the scale factor behaviour, but the matter content does
not appear directly. So, our phenomenological approach is not so
decisive in the results\cite{mnras}. Moreover, there is hopes that, due to
the
polarization of the background microwave photons, it will be possible
to measure the contribution of gravitational waves to
the anisotropy of CMBR, giving
new tests on cosmological models.
\par
For the stringlike fluid, we can find analytical solutions for the
gravitational waves, while for the cosmological constant, we obtain
analytical solutions only in the asymptotic limit.
In the case of the stringlike fluid, we can define an effective
cosmological constant $k_{eff} = k - \frac{8\pi\rho_{s0}}{3}$. We solve
the perturbed equations for $k_{eff}$ greater, lesser or equal to zero,
and we discuss the possibility of distinguish an open Universe from
a closed Universe with the dynamics of an open one. Both for the
cosmological constant and stringlike fluid case, 
we make some considerations about them as candidate for
dark matter and we analyse the implications for
the deceleration (acceleration) parameter confronting it with some
observational data.
\par
The outline of this paper is a follows: in the next section we obtain the
background solution for the two models; in the Section 3 we make the
linear
perturbative analysis in these possible Universe, and
we discuss the behaviour of gravitational waves;
some observational considerations are made in Section 4; our conclusions
are given in Section 5.

\section{Background solutions}

In order to perform a more specific analysis, we will keep ourselves in
the
simplest case:
we have two
non interacting fluids, and for each of them we define an energy-momentum
tensor which is conserved separately. These assumptions are consistent
with those of
the references quoted above. One of the energy-momentum tensor
characterizes the
ordinary matter (stiff matter, radiation, dust), and the other can
represent
a stringlike fluid or a cosmological constant.
For the case of a stringlike fluid, the spatial section can be closed,
open
or flat.
The ordinary matter has a barotropic equation of state. When the
cosmological constant
is treated, only a flat spatial section will be considered,
since this is the scenario that seems to be favored by
observations. We analyze separetely each
case.
\subsection{Stringlike fluid model}
The equations of
motion are:
\begin{eqnarray}
3(\frac{\dot a}{a})^2 + \frac{3k}{a^2} &=& 8\pi G(\rho_m + \rho_s) \quad
,\label{b1}\\
2\frac{\ddot a}{a} + (\frac{\dot a}{a})^2 + \frac{k}{a^2} &=& 
\frac{8\pi G}{3}(\rho_s - 3\alpha\rho_m)\quad ,\label{b2}\\
\dot\rho_m + 3\frac{\dot a}{a}(1 + \alpha)\rho_m &=& 0 \quad ,
\label{b3}\\
\dot\rho_s + 2\frac{\dot a}{a}\rho_s &=& 0 \quad . \label{b4}
\end{eqnarray}
In these expressions $k$ is the curvature of the spacelike section,
$\rho_m$
is the energy density of the ordinary matter,
$\rho_s$ is the energy density of the stringlike fluid, and $p_m =
\alpha\rho_m$. 
There is no direct interaction between the fluids.
Since $\rho_s \propto a^{-2}$, we can define in equation
(\ref{b1})
an effective curvature term
that can be positive, negative or zero. The
resulting equation can be written as
\begin{equation}
\frac{\dot
a^{2}}{a^{2}}-\frac{k_{eff}}{a^{2}}=\frac{\lambda}{a^{3(1+\alpha)}} \quad
,
\end{equation}
where $k_{eff}= \frac{8\pi G}{3}\rho_{0s}-k$ and $\lambda=\frac{8\pi
G}{3}\rho_{0m}$. The effective curvature term can take the 
values $k_{eff} = +\gamma$,$ -\gamma$ or $0$, where
$\gamma = \vert \frac{8\pi G}{3}\rho_{0s} - k\vert$
\par
The solutions for these equations, in terms of $k_{eff}$ and $\lambda$
and expressed in terms of the conformal time defined by $dt=ad\eta$, are
straightforward  and follow in the table below:\\\\

\begin{tabular}{||c||c|c|c||}    \hline
            & $k_{eff} > 0$                      & $k_{eff} = 0$  &
$k_{eff}
< 0$ \\ \cline{1-4}
$\alpha=-1$ &
$a=\sqrt{\frac{\gamma}{\lambda}}\quad\sin^{-1}({\sqrt{\gamma}}
~\eta)$ & $a=-(\sqrt{\lambda}~\eta)^{-1}$ &
$a=-\sqrt{\frac{\gamma}{\lambda}}\quad\sinh^{-1}(\sqrt{\gamma}~\eta)$  \\
\cline{1-4}
$\alpha=\frac{1}{3}$ &
$a=\sqrt{\frac{\lambda}{\gamma}}\sin({\sqrt{\gamma}}~\eta)$ &
$a=\sqrt{\lambda}~\eta$ &
$a=\sqrt{\frac{\lambda}{\gamma}}\sinh(\sqrt{\gamma}~\eta)$ \\ \cline{1-4} 
$\alpha=0$ &
$a=\frac{\lambda}{\gamma}\quad\sin^{2}(\frac{\sqrt{\gamma}~\eta}{2})$ &
$a=\frac{\lambda\eta^{2}}{4}$ &
$a=\frac{\lambda}{\gamma}\quad\sinh^{2}(\frac{\sqrt{\gamma}~\eta}{2})$ \\
\cline{1-4}
$\alpha=1$ &
$a=\biggr(\frac{\lambda}{\gamma}\biggl)^{\frac{1}{4}}\sqrt{\sin({2\sqrt{
\gamma}}~\eta)}$ & $a=\sqrt{2\sqrt{\lambda}~\eta}$ &
$a=\biggr(\frac{\lambda}{\gamma}\biggl)^{\frac{1}{4}}
\sqrt{\sinh(2\sqrt{\gamma}~\eta)}$ \\ \cline{1-4}
\end{tabular}
\\\\

The effective equation of state $\alpha_{eff}(\eta) =
\frac{p_T}{\rho_T}$, where $\rho_T$ and $p_T$ are the
total density and pressure respectively, changes smoothly from the
value zero (dust) or $1/3$ (radiation) to $-1/3$ (stringlike fluid). 

\subsection{Cosmological constant model}

The equations of motion, for a flat spatial section, take the form,

\begin{eqnarray}
3(\frac{\dot a}{a})^2 = 8\pi G\rho + \Lambda \quad ,\\
\dot\rho = - 3\frac{\dot a}{a}(\rho + p) \quad .
\end{eqnarray}
The pressure is, as before, related to the density as $p = \alpha\rho$,
with
$\alpha = 1$, $\frac{1}{3}$ and $0$. The solutions are
\begin{enumerate}
\item $\alpha = 1$:
\begin{equation}
a = a_0 \sinh^{1/3}(\sqrt{3\Lambda}t) \quad ;
\end{equation}
\item $\alpha = \frac{1}{3}$:
\begin{equation}
a = a_0\sinh^{1/2}(2\sqrt{\frac{\Lambda}{3}}t) \quad ;
\end{equation}
\item $\alpha = 0$:
\begin{equation}
a = a_0\sinh^{2/3}(\sqrt{\frac{3\Lambda}{4}}t) \quad .
\end{equation}
\end{enumerate}
For small values of $t$, the ordinary matter dominates and the scale
factor behaves
as in the corresponding one fluid model.
For large values of $t$, the cosmological
constant dominates, and the scale factor behaves
as in the de Sitter model, i.e.,
$a \propto e^{\sqrt{\frac{\Lambda}{3}}t}$.
As in the stringlike fluid case, the effective equation
of state $\alpha_{eff}$ evolves from $0$ (dust) or
$1/3$ (radiation) to $-1$ (cosmological constant) as the Universe expands.

\section{Evolution of gravitational waves}

The evaluation of the perturbed quantities follows
the well known approach of
Lifshitz and Khalatnikov
\cite{LK}. We will retain just the tensorial mode such that
$h_{ij} = hQ_{ij}$, where $Q_{ij}$ is a traceless transverse
eigenfunction in the three dimensional spatial section.
Perturbing the Einstein's equations, and imposing the synchronous
coordinate condition, we obtain the following equation:
\begin{equation}
\label{ss}
\ddot h - \frac{\dot a}{a}\dot h - \biggr(2\frac{\ddot a}{a}-\frac{\bar
n^{2}}{a^{2}}\biggl)h = 0 \quad ,\\
\end{equation}
where $h = \frac{h_{kk}}{a^2}$,\quad $\bar n^{2}=n^{2}+2k$,
\par
After the conformal transformation $dt=ad\eta$ in the equation (\ref{ss})
we
obtain the following equation:
\begin{equation}
\label{sss}
h'' -2\frac{a'}{a}h'+\biggr(\bar
n^{2}-2\frac{a''}{a}+2\frac{{a'}^{2}}{{a}^{2}}\biggl)h=0~~.
\end{equation}
where the primes mean derivatives with respect to $\eta$.
We will analyze now this equation, governing the evolution of
gravitational waves
in an expanding Universe, for the two configurations discussed before.
The integration of the equations follows standard procedure, and we
present the
final results only.
\subsection{Stringlike fluid}
The solutions  of the equation (\ref{sss}) for different values of the
$k_{eff}$ and different phases of the evolution of the Universe are:\\

\begin{itemize}
\item[]{\bf {1.}}\quad $k_{eff}>0$
\begin{itemize}
\item[]{\bf {(a)}}\quad $\alpha=-1$
\begin{eqnarray}
 h=\sqrt{1-x^{2}} \quad
_{2}F_{1}\biggr(2-\sqrt{1+{\tilde n}^2},2+\sqrt{1+{\tilde n}^2},
\frac{5}{2};\frac{1-x}{2}\biggl) \quad ,
\end{eqnarray}
\begin{displaymath}
\tilde n^{2}=\frac{\bar n^{2}}{\gamma}\quad , \quad\quad\quad\quad
x=\cos(\sqrt{\gamma}~\eta) \quad ;
\end{displaymath}
\end{itemize}
\begin{itemize}
\item[]{\bf {(b)}}\quad $\alpha=\frac{1}{3}$
\begin{eqnarray}
h=\exp(\mp\sqrt{1+{{\tilde n}^2}}~\eta)~~\sin({\sqrt\gamma}~\eta) \quad ,
\end{eqnarray}
\begin{displaymath}
\tilde n^{2}=\frac{\bar n^{2}}{\gamma}\quad ;
\end{displaymath}
\end{itemize}
\begin{itemize}
\item[]{\bf {(c)}}\quad $\alpha=0$
\begin{eqnarray}
h=\sqrt{1-x^{2}}~~_{2}F_{1}\biggr(-1-\sqrt{4+{\tilde
n}^2},-1+\sqrt{4+{\tilde n}^2},
-\frac{1}{2};\frac{1-x}{2}\biggl) \quad ,
\end{eqnarray}
\begin{displaymath}
\tilde n^{2}=\frac{4\bar n^{2}}{\gamma}\quad , \quad\quad\quad\quad
x=\cos(\frac{\sqrt{\gamma}~\eta}{2}) \quad ;
\end{displaymath}
\end{itemize}
\begin{itemize}
\item[]{\bf {(d)}}\quad $\alpha=1$
\begin{eqnarray}
h=\sqrt{1-x^{2}}~~_{2}F_{1}\biggr(\frac{1-\sqrt{1+4{\tilde
n}^2}}{2},\frac{1+\sqrt{1+4{\tilde n}^2}}{2},
1;\frac{1-x}{2}\biggl) \quad ,
\end{eqnarray}
\begin{displaymath}
\tilde n^{2}=\frac{\bar n^{2}}{4\gamma}\quad , \quad\quad\quad\quad
x=\cos(2\sqrt{\gamma}~\eta) \quad ;
\end{displaymath}

\end{itemize}
\end{itemize}

\begin{itemize}
\item[]{\bf {2.}}\quad $k_{eff}=0$\quad($a\propto\eta^r$)
\begin{eqnarray}
h=\eta^{\frac{2r+1}{2}}~~J_{\pm\nu}({\bar n}\eta) \quad , \quad
\nu = r + \frac{1}{2} \quad .
\end{eqnarray}
\end{itemize}
\newpage
\begin{itemize}
\item[]{\bf {3.}}\quad $k_{eff}<0$
\begin{itemize}
\item[]{\bf {(a)}}\quad $\alpha=-1$
\begin{eqnarray}
h_{1}&=&\sqrt{x^{2}-1}\biggr[\frac{x+1}{2}\biggl]^{-2+\sqrt{1-{\tilde
n}^{2}}}
\times\nonumber\\
& &{_2F_1}\biggr(2-\sqrt{1-{\tilde n}^2},\frac{1}{2}-\sqrt{1-{\tilde
n}^2},1-2\sqrt{1-{\tilde n}^{2}};\frac{2}{1+x}\biggl) \quad ,
\end{eqnarray}
\begin{displaymath}
\tilde n^{2}=\frac{\bar n^{2}}{\gamma}\quad , \quad\quad\quad\quad
x=\cos(\sqrt{\gamma}~\eta) \quad ;
\end{displaymath}\\
\begin{eqnarray}
h_{2}&=&\sqrt{x^{2}-1}\biggr[\frac{x+1}{2}\biggl]^{-2-\sqrt{1-{\tilde
n}^{2}}}\times\nonumber\\
& &{_2F_1}\biggr(\frac{1}{2}+\sqrt{1-{\tilde n}^2},2+\sqrt{1-{\tilde
n}^2},1+2\sqrt{1-{\tilde n}^{2}};\frac{2}{1+x}\biggl) \quad ,
\end{eqnarray}
\begin{displaymath}
\tilde n^{2}=\frac{\bar n^{2}}{\gamma}\quad , \quad\quad\quad\quad
x=\cos(\sqrt{\gamma}~\eta) \quad ;
\end{displaymath}\\
\end{itemize}
\begin{itemize}
\item[]{\bf {(b)}}\quad $\alpha=\frac{1}{3}$
\begin{eqnarray}
h=\exp(\pm\sqrt{1-{{\tilde n}^2}}~\eta)~~\sinh({\sqrt\gamma}~\eta) \quad ,
\end{eqnarray}
\begin{displaymath}
\tilde n^{2}=\frac{\bar n^{2}}{\gamma}\quad ;
\end{displaymath}\\
\end{itemize}
\begin{itemize}
\item[]{\bf {(c)}}\quad $\alpha=0$
\begin{eqnarray}
h_{1}&=&\sqrt{x^{2}-1}\biggr[\frac{x+1}{2}\biggl]^{-1-\sqrt{4-{\tilde
n}^{2}}}\times\nonumber\\
& &{_2F_1}\biggr(-1-\sqrt{4-{\tilde n}^2},\frac{1}{2}-\sqrt{4-{\tilde
n}^2},1-2\sqrt{4-{\tilde n}^{2}};\frac{2}{1+x}\biggl) \quad ,
\end{eqnarray}
\begin{displaymath}
\tilde n^{2}=\frac{4\bar n^{2}}{\gamma}\quad , \quad\quad\quad\quad
x=\cos(\frac{\sqrt{\gamma}~\eta}{2}) \quad ;
\end{displaymath}\\
\begin{eqnarray}
h_{2}&=&\sqrt{x^{2}-1}\biggr[\frac{x+1}{2}\biggl]^{1-\sqrt{4-{\tilde
n}^{2}}}\times\nonumber\\
& &{_2F_1}\biggr(\frac{1}{2}+\sqrt{4-{\tilde n}^2},-1+\sqrt{4-{\tilde
n}^2},1+2\sqrt{4-{\tilde n}^{2}};\frac{2}{1+x}\biggl) \quad ,
\end{eqnarray}
\begin{displaymath}
\tilde n^{2}=\frac{4\bar n^{2}}{\gamma}\quad , \quad\quad\quad\quad
x=\cos(\frac{\sqrt{\gamma}~\eta}{2}) \quad ;
\end{displaymath}
\end{itemize}
\begin{itemize}
\item[]{\bf {(d)}}\quad $\alpha=1$
\begin{eqnarray}
h_{1}&=&\sqrt{x^{2}-1}\biggr[\frac{x+1}{2}\biggl]^{\frac{-1+\sqrt{1-4{\tilde
n}^{2}}}{2}}\times\nonumber\\
& &{_2F_1}\biggr(\frac{1-\sqrt{1-4{\tilde
n}^2}}{2},\frac{1-\sqrt{1-4{\tilde
n}^2}}{2},1-\sqrt{1-4{\tilde n}^{2}};\frac{2}{1+x}\biggl) \quad ,
\end{eqnarray}
\begin{displaymath}
\tilde n^{2}=\frac{\bar n^{2}}{4\gamma}\quad , \quad\quad\quad\quad
x=\cos(2\sqrt{\gamma}~\eta) \quad ;
\end{displaymath}
\begin{eqnarray}
h_{2}&=&\sqrt{x^{2}-1}\biggr[\frac{x+1}{2}\biggl]^{\frac{-1-\sqrt{1-4{\tilde
n}^{2}}}{2}}\times\nonumber\\
& &{_2F_1}\biggr(\frac{1+\sqrt{1-4{\tilde
n}^2}}{2},\frac{1+\sqrt{1-4{\tilde
n}^2}}{2},1+\sqrt{1-4{\tilde n}^{2}};\frac{2}{1+x}\biggl) \quad ,
\end{eqnarray}
\begin{displaymath}
\tilde n^{2}=\frac{\bar n^{2}}{4\gamma}\quad , \quad\quad\quad\quad
x=\cos(2\sqrt{\gamma}~\eta) \quad ;
\end{displaymath}
\end{itemize}
\end{itemize}
where ${_2F_1}(a,b,c;x)$ are hypergeometric functions.

\subsection{Cosmological constant}

The results for the case where the cosmological constant is present are,
in the asymptotical cases, those already known in the literature
\cite{weinberg,grish}.
For small values of $t$, the ordinary fluid dominates, and we have,
\begin{equation}
h \propto t^{\frac{1}{2}(r+1)}J_{\pm\nu}(n^2 t^{1-r}) \quad , \quad \nu =
\frac{3r - 1}{2(1 -r)}
\quad .
\end{equation}
For large values of $t$, the cosmological constant dominates the matter
content of the
Universe. In this case, it is more convenient to work with the conformal
time. The
solution is:
\begin{equation}
h \propto \eta^{-1/2}J_{\pm3/2}(n\eta) \quad .
\end{equation}
We observe that, contrary to density perturbations, gravitational waves
are
produced during the deSitter phase, and in the large wavelength limit,
there is
a growing mode that evolves as $h \propto e^{2\sqrt{\frac{\Lambda}{3}}t}$.
This contrast strongly with the gravitational waves in a matter dominated
Universe, whose behaviour, in the long wavelength limit, is
\begin{equation}
h \propto t^{4/3} \quad .
\end{equation}
\section{Observational considerations}
From the solutions described above for the stringlike fluid, 
it can easily be seen that,
in what concerns the behaviour of gravitational waves, the
difference between the sign of $k_{eff}$ and $k$ itself is
negligible in the limit $n^2 \rightarrow \infty$, that is,
for small scale perturbations. The presence of the stringlike
fluid plays no significant role in this case.
However, for $n^2 \rightarrow 0$,
there are very important differences, and the sign of
$k$ plays an important role,
irrespective of the sign of $k_{eff}$. This is essential since the
measure of the anisotropy of the Cosmic Microwave Background Radiation
(CMBR) is very well established
for small values of $l$, modulus the cosmic variance problem, $l$
meaning the multipolar order in
the expansion of the two points correlation function of the temperature: 
\begin{equation}
C(\Theta) = \sum_{l=2}^\infty c_lP_l(\cos\Theta) \quad .
\end{equation}
For small $l$,
the main contribution comes from large scale perturbations, i.e.,
very small $n^2$.
\par
In order to be more precise in our statement, we will consider a specific
case in the
solutions found above. For simplicity, we take the case $\alpha =
\frac{1}{3}$ where the
solutions for the perturbation are simpler. In all other cases, however,
the
reasoning is
the same. Taking $k_{eff} < 0$ in the limit $n^2 \rightarrow 0$, we find,
\begin{equation}
h \propto e^{\pm\sqrt{1-2k}~\eta}\sinh\sqrt{\gamma}~\eta \quad .
\end{equation}
Hence, we obtain the following expressions in function of the sign of $k$:
\begin{itemize}
\item $k = - 1$ (open Universe):
\begin{equation}
h_\pm \propto e^{\pm\sqrt{3}\eta}\sinh\sqrt{\gamma}~\eta \quad ;
\end{equation}
\item $k = 0$ (flat Universe):
\begin{equation}
h_\pm \propto e^{\pm\eta}\sinh\sqrt{\gamma}~\eta \quad ;
\end{equation}
\item $k = 1$ (closed Universe)
\begin{equation}
h \propto \cos\eta\sinh\sqrt{\gamma}~\eta \quad .
\end{equation}
\end{itemize}
\noindent
In the same limit, for $k_{eff} = 0$ 
we get the folowing expression,
\begin{itemize}
\item $k = 1$ (closed Universe)
\begin{equation}
h \propto  \eta^{2r+1} \quad .
\end{equation}
\end{itemize}
while for $k_{eff} > 0$, we get
\begin{itemize}
\item $k = 1$ (closed Universe)
\begin{equation}
h \propto e^{\pm\sqrt{3}\eta}\sin\sqrt{\gamma}~\eta \quad .
\end{equation}
\end{itemize}
Note that $k = 1$ admits all three possible values for $k_{eff}$,
while $k =0$,$- 1$ lead to $k_{eff} < 0$.
\par
The relevant observable quantity, the two points correlation function
of the fractional fluctuation in the observed background temperature,
has an expression that depends strongly on the seeds of the perturbations,
and on the behaviour of perturbed quantitites, like $h$. If we take $k =
1$,
and $k_{eff} < 0$,
the behaviour of $h$ has features completely different
with respect to an open Universe. Hence, in principle, a closed
Universe with a dynamics of an open one can be tested by the
observation.
\par
In the cosmological constant model, we have already seen that
in the long wavelength limit, there is a very clear difference
between the behaviour of gravitational waves in the matter dominated
era and in cosmological constant dominated era. This must reflects
in the anisotropy of CMBR provocated by a cosmological constant.
We remark that, in this respect, this behaviour of gravitational wave
in presence of a cosmological constant is clearly distinct from the
the behaviour of density perturbations: density perturbations
generated by a cosmological constant are zero, so the determinant role
is played by the ordinary fluid.
\par
In all these cases, we must observe that the two point correlation
function depends also on the geometry of the three dimensional
spatial section. The eigenfunctions $Q_{ij}$ are of course not
the same if the spatial section is flat, closed or open.
\par
The existence of a stringlike fluid or a cosmological constant
may be reflected in the value of total density of the Universe,
$\Omega_T$. The observational determination of $\Omega$ 
remains an open problem
in cosmology\cite{ellis}. If the limits
coming from the primordial nucleosynthesis
are taking into account, the baryonic mass parameter
is $\Omega_B \sim 0.02$. However, the dynamics of galaxie cluster leads
to $\Omega \sim 0.3$. Moreover, the doppler peaks present in the
$c_l$ spectrum for the anisotropy of CMBR seems to be consistent with
$\Omega_T \sim 1$. More recently, deviation of linearity of Hubble's law
may suggest a flat Universe that is accelerating. If this result is
confirmed, this is a strong evidence in favor of the existence
of a cosmological constant. We remark however that the stringlike fluid
may account for a fraction of dark matter of order $\Omega_s = 0.7$
only if it is a representation of a variable cosmological constant.
If it represents a fluid of cosmic string, it will contribute for
the clustered mass only\cite{sergio}
\par
Indeed, observations of supernova in the
redshif range $0.16 < z < 0.62$ favours an accelerating Universe.
What is the consequence of that for our models?
The deccelerating parameter is given by $q_0 = - \frac{\ddot a a}{\dot
a^2}$.
We apply this expression for our two models above, for the case
$\alpha = 0$, since the observations are made today.
\begin{itemize}
\item Cosmological constant model:
\begin{equation}
q_0 = 2 - \frac{3}{2}\tan^2\sqrt{\frac{3\Lambda}{4}t} \quad ;
\end{equation}
\item Stringlike fluid model ($k_{eff} < 0$):
\begin{equation}
\label{qcs}
q_0 = \frac{1}{2}\frac{1}{\cosh^2\frac{\sqrt{\lambda}\eta}{2}} \quad.
\end{equation}
\end{itemize}
In the cosmological constant case, the Universe is initially 
decelerating, and from a time defined by
\begin{equation}
t_c = \sqrt{\frac{4}{3\Lambda}}\tanh^{-1}\frac{4}{9}
\end{equation}
it begins to be accelerated.
However for a stringlike fluid, the Universe is always decelerating.
Hence, the confirmation of the results coming from the supernova
sample may lead to discard the stringlike phenomenological model
considered here, unless the observational data allow
$q_0 \sim 0$ today, which is the asymptotic limit for (\ref{qcs}).
\section{Conclusions}
In this article, we have discussed the evolution of the gravitational
waves in the
two-fluid models, consisting in the ordinary matter and the exotic matter
whose equation of state is $p=-\rho/3$ 
(stringlike fluid) or $p = - \rho$
(cosmological constant). For the first case,
we can define an effective curvature parameter $k_{eff} = k - \lambda$,
where $\lambda$ is linked to the stringlike fluid density.
The present study applies for all
values of $k_{eff}$, generalizing the results of the preceding
work that only treated the case $k=1$\cite{mnras}.
In the second case, we have considered just a spatially flat Universe.
\par
The solution of the linear perturbed equation, 
for the stringlike fluid configuration, is expressed in terms
of hypergeometric functions. It comes out that the behaviour of
this model, concerning
gravitational waves, is strongly depending not only on the value of
$k_{eff}$
but also on the value of $k$. In particular for $k=1$ the behaviour
of gravitational wave is completely different if
$k_{eff}=-1$, $0$ or $1$. For $k = 0$ and $k = -1$, we have
necessarily $k_{eff} < 0$, and only in the case $k = - 1$ the
behaviour of gravitational wave is essentially the same as
in the open Universe with no stringlike fluid.
We remark that the scale factor behaviour of the background does
not permit to distinguish between the sign of $k$ and $k_{eff}$.
But this is not the case for the gravitational waves.
The most important case is when $k = 1$ and $k_{eff} < 0$, that is
a closed Universe exhibiting the behaviour of an open one.
Here, in the long wavelength limit, the gravitational waves behave
in a complete different way with respect to an open Universe in a one
fluid approach: in the last case, we have growing modes, while for
the former one the amplitude of gravitational waves oscillates.
In what concerns the spectrum of anisotropy of the CMBR,
it is possible to distinguish all possible combinations
of sign of $k$ and $k_{eff}$, except $k = - 1$ and
$k_{eff} < 0$, due to
the different expansion into harmonic functions.
\par
For the cosmological constant model, the behaviour of gravitational
waves has specific features which may permit to distinguish it from
a one fluid model with ordinary matter. In particular, in the long
wavelength limit the gravitational waves are strongly amplified
when we enter in a phase where the cosmological constant dominates.
The existence of a cosmological constant can be reflected,
for example, in the position of the first doppler peak
in the CMBR anisotropy spectrum, since it depends on the inverse
of the square of the total mass; however, the position of the first
doppler
peak may indicate the existence of a dark matter, but does not reveal
in principle its nature. It can be, for example, a stringlike
fluid as considered here or some other exotic fluid. But, depending
on the fundamental field the stringlike fluid represents, it
can contribute for the clustered or unclustered mass.
\par
Recently, however, it has been argued that analysis of a sample of
supernova reveals a deviation of Hubble's law from linearity
that is consistent with a cosmological model with $\Omega_T \sim 1$
and $\Omega_\Lambda = 0.7$\cite{riess}. This analysis seems to show that
the
Universe is an accelerating phase. If this is the case, the
stringlike fluid model considered here may be disregard,
since it predicts $q_0 > 0$ (decelerating Universe) unless $q_0 \sim 0$
is also allowed, which is its asymptotic limit.
For the cosmological constant model, there is an initial
phase for which $q_0$ is positive, then negative from
a transition time $t_c$ on. We note that an accelerating
Universe would be
a very strong indication of the existence of dark matter whose equation of
state is such that $p < - \frac{\rho}{3}$ (since
this equation of state implies a violation of the strong energy
condition and consequently leads to an accelerated Universe),
the stringlike
fluid considered here
being a lower limit and the cosmological constant the
most natural candidate\cite{jerome}.
\par
In order to have a better comparison with observations,
we should calculate the spectrum of perturbations and
the coefficients $c_l$ related to the anisotropy of CMBR.
This has been done for example for the case where the exotic
fluid is the so called quintessence or a variable cosmological
constant\cite{quinte2,coble}. However, to
do so, we should first evaluate density perturbations and its
corresponding transfer function, and this lies outside the scope of
the present work.

\section*{Acknowledgements}
 It is a pleasure to thank
 J\'er\^ome Martin and Marco Picco for many usefull discussions.
 We thank CNPq and CAPES (Brazil) for financial support of this work.
 J.C.F. would like to thank the hospitality of the {\it Laboratoire de
 Gravitation et Cosmologie Relativistes}, University of Paris VI,
 during the elaboration of this work.

\end{document}